\documentclass[fleqn,twoside]{article}
\usepackage{espcrc2_modified}

\usepackage{graphicx}
\usepackage[figuresright]{rotating}

\newcommand{\AmS}{{\protect\the\textfont2
  A\kern-.1667em\lower.5ex\hbox{M}\kern-.125emS}}

\title{Status of MiniBooNE}

\author{Andrew O. Bazarko\address[]{Department of Physics, 
Princeton University, Princeton, New Jersey 08544-0708, USA}
for the BooNE Collaboration
\thanks{To appear in Proceedings of the 31st International Conference on
High Energy Physics (ICHEP02) Amsterdam, 24-31 July 2002.}
} 
      
\begin{document}

\begin{abstract}
MiniBooNE is a neutrino oscillation experiment now running 
at Fermilab.  The experiment will search for $\nu_\mu\to\nu_e$
oscillations in order to make a conclusive statement about 
the yet-unconfirmed evidence for oscillations presented by the 
LSND experiment.  Preparations for the start of running 
were completed over the summer of 2002, and MiniBooNE observed
its first neutrino events in late August.  
\vspace{1pc}
\end{abstract}

\maketitle

\section{LSND EVIDENCE}

The Liquid Scintillator Neutrino Detector ran at the Los Alamos 
Neutron Science Center over the years 1993 to 1998.
Using the accelerator's 800 MeV proton beam, the experiment
studied neutrinos coming from the decay  $\mu^+\to e^+\nu_e\bar\nu_\mu$
of muons at rest and from the decay $\pi^+\to\mu^+\nu_\mu$ of pions
in flight.  The muon decay at rest (DAR) process allowed a search
for $\bar\nu_\mu\to\bar\nu_e$ appearance via the reaction 
$\bar\nu_e p \to e^+ n$;
neutrinos from pion decay in flight (DIF) provided
a search for $\nu_\mu\to\nu_e$ appearance via the reaction
$\nu_e C\to e^- N$.

The LSND detector was a cylindrical tank of
167 tons of mineral oil doped with a small amount of scintillator 
viewed by 1220 photomultiplier tubes. 
It was positioned approximately 30 m from 
the beam target area.  For the 1993-95 running period the primary proton 
target was a water target, and in 1996-98 the target was made mostly of 
tungsten.  
The neutrino yield from the 
water target was higher, so that about 59\% of the total 
DAR flux and 62\% of the total DIF flux came from the 1993-95 period. 

The DAR process yields neutrinos with energies up to the 52.8 MeV 
decay endpoint, and the DIF process produced $\nu_\mu$ up to 300 MeV.   
The detector uses Cherenkov light for electron particle 
identification.  
The reaction $\bar\nu_e p \to e^+ n$
is observed by correlating the electron track 
with the 2.2 MeV gamma ray that follows from neutron
capture on free protons in the liquid, $n p \to d \gamma$.

\begin{figure}[tbh]
\vspace{-12pt}
\includegraphics[width=75mm]{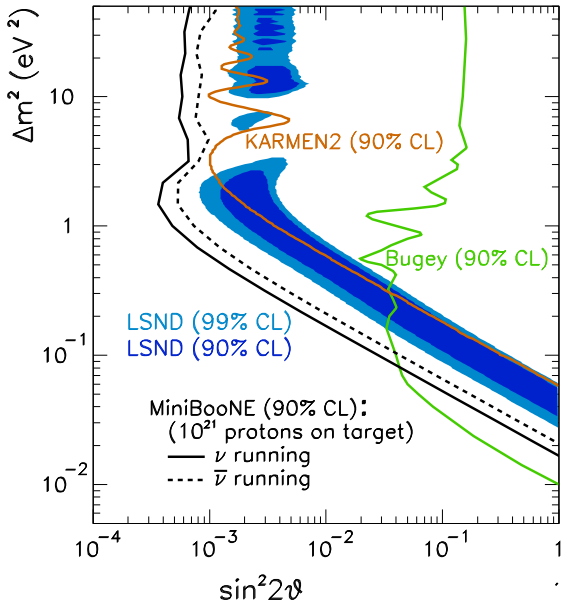}
\vspace*{-30pt}
\caption{The LSND 90\% and 99\% C.L. allowed regions compared 
with the 90\% C.L. exclusion contours from KARMEN 2 
$(\bar\nu_\mu\to\bar\nu_e)$, Bugey $(\bar\nu_e\to\bar\nu_\mu)$,
and the expected MiniBooNE contours using a $\nu_\mu$ or a 
$\bar\nu_\mu$ beam.} 
\label{fig:l_k}
\vspace*{-4mm}
\end{figure}

LSND has presented results periodically.
Analysis of 1993-95 DAR data  
produced an excess above backgrounds of $51.0^{+20.2}_{-19.5}\pm8.0$ 
events, corresponding to an oscillation probability of 
$(0.31\pm0.13)$\% (combined statistical and 
systematic errors) \cite{lsnd96}.  The DIF data
from the same period yielded an excess of $18.1\pm6.6\pm4.0$
events, and an oscillation probability of 
$(0.26\pm0.11)$\% \cite{lsnd98}.  

In 2001, LSND presented final results using all of its
data, combining the $\bar\nu_\mu\to\bar\nu_e$ and $\nu_\mu\to\nu_e$
searches into a single analysis,
and employing new event reconstruction with better spatial 
resolution \cite{lsnd01}.  
A total excess of $87.9\pm22.4\pm6.0$ events was observed in the
$\bar\nu_\mu\to\bar\nu_e$ search, corresponding to an oscillation 
probability of $(0.264\pm0.081)$\%. 
The common event selection was optimized
for the DAR region below 60 MeV, and was therefore less effective 
than the previous DIF analysis in removing background events above 60 MeV. 
This analysis found 
no significant signal 
in the DIF energy region, where the observed excess was 
$8.1\pm12.2\pm1.7$ events.

The Karlsruhe Rutherford Medium Energy Neutrino experiment at the 
ISIS facility of the Rutherford Lab also 
searched for $\bar\nu_\mu\to\bar\nu_e$ 
oscillations using $\bar\nu_\mu$ from $\mu^+$ decay at rest \cite{karmen2}.  
The experiment employed a 56 ton segmented liquid scintillator detector 
located about 18 m from the beam target. The 1997-2001 run, known at 
KARMEN 2, did not observe an excess of events, finding 15 events while 
expecting a background of $15.8\pm0.5$ events.  However, KARMEN 2 only 
had the sensitivity to rule out a portion of the LSND allowed parameter 
space.  A comparison of the favored LSND oscillation parameters with the 
KARMEN 2 limit is shown in Fig. \ref{fig:l_k}. 
A combined statistical analysis of LSND and KARMEN 2  
finds regions of oscillation parameters compatible with both experiments \cite{lk_comb}.


The LSND neutrino mass difference, $\Delta m^2_{LSND}> 0.1$ eV$^2$, is 
distinct from those indicated by oscillations of solar and atmospheric 
neutrinos \cite{giunti}. 
This poses the problem that three neutrino masses cannot explain the 
three $\Delta m^2$'s.  Proposed solutions include consideration of a 
sterile neutrino, supersymmetry, $CPT$ violation, or lepton number 
violating muon decay \cite{babu}.  
In addition, the relatively high neutrino mass would 
contribute to the question of dark matter.  It remains important to 
confirm or refute the LSND evidence, which MiniBooNE sets out to do.

\section{MINIBOONE OVERVIEW}

At MiniBooNE we have begun collecting data to search for 
$\nu_\mu \rightarrow \nu_e$ oscillations.  
The $\nu_\mu$ beam energy is in the range
0.5 -- 1.5 GeV with a small intrinsic
$\nu_e$ component and we will search for an excess of
electron neutrino events in a detector located approximately 500 m from the neutrino source.  
The baseline to neutrino energy ratio is thereby similar to that of 
LSND, $L/E \sim 1$ m/MeV, while neutrino energies are more than an order 
of magnitude higher.

The MiniBooNE neutrino beam is initiated by a primary beam of 8 GeV protons
from the Fermilab Booster.  The Booster is a reliable, high intensity 
machine, expected to run at least $2\times 10^7$ s per year, 
while delivering $5\times 10^{12}$ protons per 1.6 $\mu$s pulse
at a rate of 5 Hz to MiniBooNE.  By this proton accounting, 
one nominal year for 
the experiment is expected to supply $5\times10^{20}$ protons on target.
The Booster has the capacity to provide protons for  
several Fermilab efforts, and the Booster
continues to deliver protons to the TeVatron collider program.

\begin{figure}[tb]
\vspace{9pt}
\includegraphics[width=65mm]{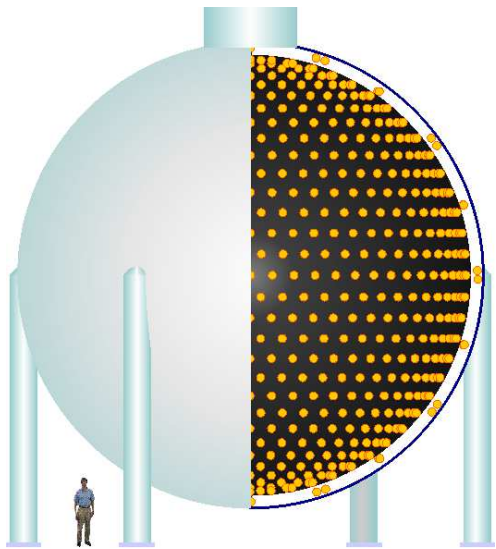}
\caption{A schematic of the MiniBooNE detector.  The cutaway shows the layout of 
20-cm phototubes in the black main region and in the white veto region.} 
\label{fig:detector}
\vspace*{-4mm}
\end{figure}

A secondary beam is produced when 
the 8 GeV protons strike a beryllium target positioned inside  
a magnetic horn.  
At present, positively charged particles (mostly $\pi^+$'s) from  
the target are focused forward by the single horn into a 50 m decay tunnel.
The $\nu_\mu$ beam is produced from the decay of these secondary particles. 
At a later time, the polarity of the horn can be changed in order to 
focus $\pi^-$'s to generate a $\bar\nu_\mu$ beam. 
Decay lengths of 50 and 25 m are possible through the use of 
two steel and concrete beam absorbers.  
One absorber is permanently positioned at the end of the decay tunnel.
The intermediate absorber can either be lowered into the 
decay tunnel or raised out of the way.  
This ability to vary the decay length will provide a check of experimental 
systematics associated with $\nu_e$ contamination.

\begin{figure}[bt]
\vspace{9pt}
\includegraphics[width=72mm]{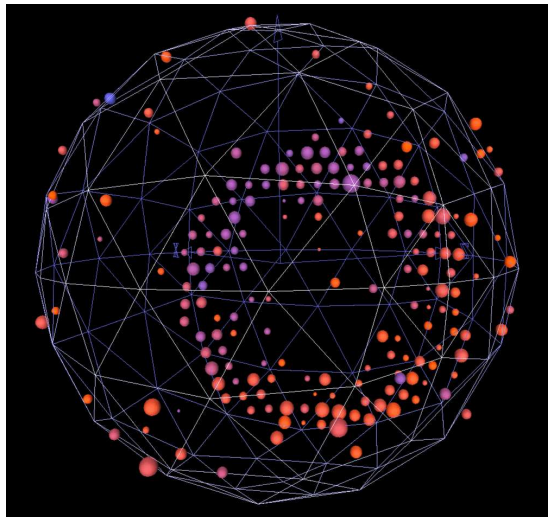}
\caption{Event display of the muon signature from a candidate $\nu_\mu$ interaction.  
Each dot represents 
a hit photomultiplier tube with the size proportional to the charge collected.  
This event comes from the set
of the first few beam events observed by MiniBooNE in late August 2002.}
\label{fig:event}
\vspace*{-4mm}
\end{figure}

The MiniBooNE neutrino 
detector consists of 800 tons of pure mineral oil contained in a 40-foot (12.2 m) 
diameter spherical tank.  A structure in the tank supports phototubes
and optically isolates the most outer 35 cm 
of oil from the rest, turning the outer oil into a veto region that should stay quiet
while a neutrino produces light only in the inner, main region.
The main region is viewed by 1280 20-cm phototubes, providing
10\% photocathode coverage of the 445 ton fiducial volume.
The veto region contains 240 20-cm phototubes mounted in pairs on the tank wall.  
A schematic of the detector is shown in Fig. \ref{fig:detector}.
The center of the detector is positioned about 490 m from 
the end of the decay tunnel, and about 6 m below
ground, corresponding to the level of the neutrino beam.  
An overburden of about 3 m of soil shields the detector enclosure. 

The detector records the arrival time and total charge 
for each hit phototube.   From this information, the track position
and direction are determined.  Electrons from $\nu_e$ 
interactions are identified via their characteristic
Cherenkov and scintillation light signatures (the undoped oil scintillates modestly). 
Backgrounds to the oscillation search 
will be due to  $\nu_e$ contamination in the beam
and to the misidentification as electrons of 
muons and $\pi^0$'s produced in the detector.  
A display of one of MiniBooNE's first candidate neutrino events is 
shown in Fig. \ref{fig:event}. 


\section{BEAMLINE STATUS}

Intense activity over summer 2002 focused on preparing the new 8 GeV proton 
beamline coming off of the Fermilab Booster to the MiniBooNE production 
target and secondary beam focusing horn.  
MiniBooNE started receiving proton beam on target in late August. 

The focusing horn was designed to run at 5 Hz and for 200 Mpulses --- a 
higher pulse rate and more pulses than any previous horn.  In fall 2001 the horn 
was successfully bench-tested with more than 10 Mpulses
making it already the world's most pulsed horn. 

In summer 2002 the proton beam was commissioned in a few stages.  In June protons were sent 
through the new beamline to a stop ahead of the MiniBooNE target area.  Next the target and 
horn were installed and then removed in a dry run of procedures to be used for removing a 
damaged, radioactive horn.  In July, without the target and horn in place, protons were sent 
through the target location, where a temporary multiwire beam monitor recorded the beam 
profile.  The multiwire measurements were used to commission the final focus and to begin 
calibration of permanent beam position monitors (BPM's).
In August the target and horn were reinstalled.
The final configuration was modified slightly to include a multiwire just 
upstream of the target, which was used to complete the calibration of the 
target BPM's.


High intensity Booster operation is restricted by radiation limits, 
and beam losses will have to be reduced before MiniBooNE can reach its full beam intensity. 
In September the Booster was typically supplying $4\times 10^{12}$ protons per pulse. 
(Losses were observed to grow disproportionately when trying to run with $5\times 10^{12}$ 
protons per pulse.)  By mid-September the pulses were arriving on target at 1 Hz. 
The experimental priority is to increase the repetition rate to 5 Hz (whereas $4\times 10^{12}$
protons per pulse is adequate).  
Operation above 1 Hz will become possible with the planned installation of a new extraction 
magnet in the winter.   However, under current run conditions 5 Hz operation would exceed
the radiation limits for beamline elements.  


The Booster now supplies protons to both the TeVatron collider and to  
MiniBooNE.  The start of MiniBooNE running has had negligible effect on collider operation. 

\section{DETECTOR STATUS} 

Installation of phototubes and of all related in-tank calibration and monitoring 
hardware was completed in September 2001.  Filling the tank with oil started in 
December and completed in May 2002.  The 
data acquisition electronics were commissioned over the same period, and phototube
signals produced by cosmic rays and by a laser calibration system indicated the 
rising oil level.  
 
The laser calibration system provides pulses 
to four light-diffusing flasks at various locations in the detector.  The system is being 
used to determine phototube time offsets and time slewing
corrections.  Energy calibrations are being performed using 
electrons from the decay of stopped cosmic ray muons.
In addition, a cosmic ray calibration system is being commissioned, which 
consists of a hodoscope muon tracker
above the detector with seven scintillator cubes (7.6 and 10  cm on a side)  
located under the tracker inside the detector.  The system will provide 
the entering position and direction of cosmic ray muons.
Cosmic rays that stop in a cube are a sample of muons with well-known path length 
in the detector, and will be used to calibrate the position, energy, and 
direction determination of the reconstruction algorithms.

\section{PROSPECTS}

As the first neutrinos were delivered to MiniBooNE the detector was ready 
to record them.  It now remains to be seen how reliable and intense 
the beam will be.  The beam intensity came up quickly, and within about three
weeks of the start of the run, $4\times 10^{12}$ protons per pulse at
a rate of 1 Hz were being delivered.  The 1 Hz operation is currently limited
by beamline hardware, which will be replaced this winter to allow 5 Hz operation.
Work is ongoing to control beam losses so that 5 Hz beam operation is possible
within radiation limits.

The experiment is currently running with a $\nu_\mu$ beam.  If oscillations occur 
as indicated by LSND, with $10^{21}$ protons on target MiniBooNE will observe an
excess of several hundred electron events that will be significantly above background.   
The option will exist during the run to reverse 
the horn polarity to provide a $\bar\nu_\mu$ beam.  The  
$\bar\nu_\mu$ beam will have a reduced flux, and the antineutrino cross section is smaller,   
but the backgrounds are also expected to be smaller, so that the experiment 
would retain significant sensitivity.  MiniBooNE's expected 90\% C.L. exclusion contours
using a $\nu_\mu$ and a $\bar\nu_\mu$ beam are shown in Fig. \ref{fig:l_k}.


\begin{thebibliography}{9}

\bibitem{lsnd96} LSND, Phys. Rev. C {\bf 54} (1996) 2685.
\bibitem{lsnd98} LSND, Phys. Rev. C {\bf 58} (1998) 2489.
\bibitem{lsnd01} LSND, Phys. Rev. D {\bf 64} (2001) 112007.
\bibitem{karmen2} KARMEN, Phys. Rev. D {\bf 65} (2002) 112001.
\bibitem{lk_comb} E.D. Church, et al. Phys. Rev. D {\bf 66} (2002) 013001.
\bibitem{giunti} C. Giunti, these proceedings.
\bibitem{babu} K.S. Babu, these proceedings.

\end{thebibliography}
\end{document}